# Calibration-Free Cuffless Blood Pressure Estimation Using Multimodal ECG-PPG Fusion on a Google Pixel Watch

Jathushan Kaetheeswaran

Institute of Biomedical Engineering, University of Toronto, Canada and KITE Research Institute, University Health Network, Canada

Boyi Ma

Institute of Biomedical Engineering, University of Toronto, Canada and KITE Research Institute, University Health Network, Canada

Ali Abedi

Peter Munk Cardiac Centre, University Health Network, Canada

Shehroz S. Khan*

College of Engineering and Technology, American University of the Middle East, Egaila, 54200, Kuwait

Milad Lankarany*

Institute of Biomedical Engineering, University of Toronto, Canada and KITE Research Institute, University Health Network, Canada

* These authors contributed equally to this work

Home blood pressure (BP) monitoring is a critical component of long-term cardiovascular management to identify and reduce cardiovascular risks. However, existing cuff-based gold-standard devices are limited by infrequent measurements, restrictive operating conditions, and sensitivity to posture and motion. Consumer-grade wrist-worn devices such as smartwatches provide an opportunity for continuous BP estimation using multimodal physiological sensing. Due to the absence of an inflatable cuff, consumer-grade smartwatches requires robust algorithms to estimate BP from relevant biosignals. This work develops and benchmarks mechanism-based, machine learning, and deep learning (DL) approaches for cuffless BP estimation on a Google Pixel Watch using the HEART-Watch dataset, a synchronized Google Pixel Watch dataset containing ECG, PPG, and cuff-based BP measurements from 40 participants. A multimodal DL architecture was developed to use raw 8-second smartwatch ECG and PPG inputs, modulate learned signal representations using demographic encodings, and estimate systolic and diastolic BP. Under a calibration-free leave-one-subject-out cross-validation framework, the proposed model achieved mean absolute errors of 9.13 mmHg for systolic BP and 7.77 mmHg for diastolic BP, demonstrating competitive calibration-free performance on consumer-grade smartwatches. Ablation analyses demonstrated that multimodal fusion of ECG, PPG, and demographic inputs improved generalizability to unseen subjects compared to reduced-input models. Subgroup analysis further revealed statistically significant increases in diastolic BP estimation error for obese participants compared to non-obese participants, suggesting that equitable model performance is dependent on capturing inter-subject physiology through model inputs. Overall, our findings demonstrate the potential of consumer-grade smartwatches for calibration-free cuffless BP estimation while emphasizing the need for larger datasets, additional physiological sensing modalities, and advanced fusion techniques to improve reliability for real-world deployment.

## AUTHOR SUMMARY

Inadequate blood pressure (BP) monitoring and management outside of clinical settings can worsen major cardiovascular risk factors such as hypertension. While cuff-based devices are commonly used for at-home monitoring, these devices can be inconvenient for daily use due to their sensitivity to body positions, upper-arm constrictions, and limited portability. A promising alternative is emerging in the form of consumer-grade smartwatches, where physiological signals related to cardiac activity can be used to estimate BP non-invasively and continuously across daily living conditions. In this work, we use data collected from a Google Pixel Watch in 40 participants to develop and compare several algorithm approaches for BP estimation. We found that our proposed deep learning model achieved the strongest overall performance, and that fusing smartwatch signals with demographic information improved model generalizability to unseen individuals. However, we also identified that model accuracy was not consistent across participant subgroups, with obese individuals yielding higher estimation errors than others. This study highlights the feasibility of consumer-grade smartwatches as accessible platforms for deploying robust BP estimation algorithms, though clinical reliability will require larger, more diverse populations and additional sensing modalities.

## 1 INTRODUCTION

Hypertension remains one of the leading modifiable predictors of cardiovascular disease [51], with long term blood pressure (BP) trends strongly predicting cardiovascular health outcomes [53]. A longitudinal study (1996–2021) found hypertensive children were more than twice as likely to experience major cardiac events in adulthood [54]. Timely interventions and cardiovascular management are therefore strongly dependent on the quality of tools available for longitudinal BP monitoring [44]. Continuous self-monitoring of BP, when paired with telemedicine, has been associated with improved health outcomes [7]. Clinical measurements of BP are regularly conducted during healthcare checkups; however, due to their limited frequency and constant setting, they may fail to capture patterns such as masked and white-coat hypertension [51, 68]. Masked hypertension refers to cases where clinical measurements indicate normotensive readings, while home BP monitoring (HBPM) falls within hypertensive ranges. White-coat hypertension describes the opposite phenomenon, where HBPM readings remain within normal ranges, but clinical measurements obtained in controlled environments are elevated. Despite the growing need for HBPM technologies to capture these natural variations in BP, accessible long-term monitoring solutions remain limited.

Traditional cuff-based sphygmomanometers are the current gold-standard for HBPM; however, they are cumbersome for daily use and provide only snapshot readings [61, 72]. These devices are often automatic and susceptible to inaccuracies due to body position, motion, and improper cuff dimensions, all of which are common in uncontrolled home environments [51]. In contrast, wrist-worn wearables are rapidly emerging as multimodal alternatives equipped with electrocardiogram (ECG) and photoplethysmography (PPG) sensors for continuous HBPM. In addition to housing the necessary sensing hardware for non-invasive BP estimation, consumer-grade wrist-worn devices such as smartwatches are widely adopted due to their portability and flexibility as both healthcare and lifestyle tools [17]. The growing commercial interest in wearable HBPM is exemplified by the Apple Watch recently obtaining FDA clearance for an AI-based hypertension notification feature using PPG [6, 34]. Although several cuffless BP watches (e.g., Heartisans, Omron HeartGuide) are commercially available, many demonstrate poor agreement with traditional cuff-based measurements, partly due to limited global validation standards [16, 35, 50].

Robust BP estimation using wrist-worn wearables remains a challenging research problem. The mappings between wrist-worn biosignals and BP are influenced by confounding factors such as inter-subject arterial stiffness and physical activity, introducing non-linearities into the estimation process [73]. In addition to these physiological challenges, wearable technologies are constrained by signal contaminants associated with uncontrolled daily use, including poor wrist contact and motion artifacts. Table 1 provides a non-exhaustive summary of previous research works for cuffless BP estimation using wrist-worn technologies. The majority of studies rely on custom-built devices for data collection and validation. Although these works demonstrate the potential of wrist-worn algorithms for HBPM, they may not accurately reflect real-world performance when deployed on existing consumer-grade hardware. Furthermore, many existing research works do not employ fully wrist-based multimodal strategies, instead relying on PPG-only approaches and/or additional instrumentation such as chest ECG systems. Studies that do employ wrist-based ECGs are either custom/prototype devices [21, 23, 32] or dedicated fitness tracking bands [40, 61], which cannot be fairly compared with the sensing constraints and signal characteristics of mass-produced smartwatches. Dedicated devices often incorporate high-fidelity sensors and specialized hardware designs that can inflate BP estimation performance under controlled conditions due to higher signal quality [19]. In contrast, commercial smartwatches are designed as general-purpose consumer devices rather than specialized cardiovascular monitoring systems. Factors such as industrial design constraints, user comfort, fixed sensing configurations, and onboard processing limitations can influence biosignal quality and downstream BP estimation performance differently from custom research hardware [24, 55].

This work investigates how multimodal signals acquired directly from a consumer-grade smartwatch (Google Pixel Watch) can be utilized for BP estimation. To the best of our knowledge, the HEART-Watch dataset is the first dataset to provide synchronized ECG and PPG signals from a commercially available smartwatch alongside cuff BP measurements. Unlike previous studies focused on specialized hardware development, this work evaluates the feasibility of cuffless BP estimation under the sensing constraints of a consumer-grade smartwatch during semi-naturalistic conditions.

Table 1. Non-exhaustive summary of previous algorithm development for cuffless BP estimation using wrist-worn wearable ECG, PPG, and/or ACC signals.

| Reference | Subjects (F/M) | Wearable device | Biosignals | Physical states | BP Reference | # of BP observations | Model | Window size | MAE ± STD (mmHg) |
|---|---|---|---|---|---|---|---|---|---|
| Thomas et al. 2014 [63] | 4 (—) | Custom watch | ECG, PPG | Supine, Sitting, Standing | Continuous | — | PWV models | 1 beat | — |
| Hsiao et al. 2016 [23] | 120 (54/66) | Custom wristband | ECG, PPG | Sitting | Snapshot | 240 | HR-MK, HR-A-MK, HR-BMI-MK, HR-A-BMI-MK | 1 minute | SBP: 6.72 ± 8.50<br>DBP: — |
| Atomi et al. 2017 [3] | Training dataset: — (—)<br><br>Testing dataset: 25 (—) | Denso Corporation watch | PPG | Sitting | Training dataset: Snapshot<br><br>Testing dataset: Continuous | Training dataset: 632<br><br>Testing dataset: 122 | Multiple LR | — | SBP: — ± 8.54<br>DBP: — |

| Reference | Subjects (F/M) | Wearable device | Biosignals | Physical states | BP Reference | # of BP observations | Model | Window size | MAE ± STD (mmHg) |
|---|---|---|---|---|---|---|---|---|---|
| Carek et al. 2017 [9] | 13 (—) | Custom watch | ACC, PPG | Sitting, Stair stepping | Continuous | — | R-MK | 10 beats | SBP: —<br>DBP: 5.0 ± 5.0 |
| Lazazzera et al. 2019 [27] | LDB dataset: 5 (0/5)<br><br>TDB dataset: 44 (—) | Custom watch | PPG | Sitting | LDB dataset: Snapshot<br><br>TDB dataset: Snapshot | LDB dataset: 22<br><br>TDB dataset: 122 | HR-MK | 30 seconds | SBP: — ± 9.45<br>DBP: — ± 4.93 |
| Mena et al. 2020 [37] | 3 (2/1) | Custom wristband | PPG | Sitting | Snapshot | 150 | ANN | ≤ 30 seconds | SBP: — ± 8.58<br>DBP: — ± 4.21 |
| Moon et al. 2020 [40] | 35 (18/17) | InBody watch | ECG, PPG | Sitting | Snapshot | 245 | FNN | 24 seconds | SBP: — ± 7.3<br>DBP: — ± 4.2 |
| Song et al. 2020 [61] | Training dataset: 66 (41/25)<br><br>Testing dataset: 44 (34/10) | InBody watch | ECG, PPG | Sitting | Snapshot | ≥ 330 | Stacked DNN | 20 seconds | SBP: 4.8 ± 6.0<br>DBP: 4.8 ± 6.0 |
| Cao et al. 2021 [8] | 35 (18/17) | Custom wristband | PPG | Sitting | Snapshot | — | BLSTM-HNN | 1 beat | SBP: — ± 6.55<br>DBP: — ± 7.31 |
| Ganti et al. 2021 [19] | 21 (5/16) | Custom watch | ACC, PPG | At home | Snapshot | ≥ 245 | R-MK | 15 seconds | SBP: 4.03 ± 2.29<br>DBP: 2.24 ± 0.75 |
| Paliakaite et al. 2021 [43] | 22 (11/11) | Custom wristband | PPG | Sitting, Cold pressor | Continuous | — | Subject-specific LR | 1 beat | SBP: 9.69 ± 12.86<br>DBP: 7.76 ± 10.20 |
| Pediaditis et al. 2021 | 11 (—) | Custom watch | PPG | Sitting | Snapshot | ≥ 88 | RF, SVR, MLP | 10 seconds | SBP: 7.65 ± —<br>DBP: 9.45 ± — |
| He et al. 2022 [21] | 76 (34/42) | Custom watch | ECG, PPG | Sitting | Snapshot | 2726 (1 label for 3min 20 seconds) | LR & TrAdaBoost | 5 seconds | SBP: 7.05 ± 9.36<br>DBP: 5.79 ± 7.02 |
| Yao et al. 2022 [71] | 33 (12/21) | Custom watch | PPG | Sitting | Snapshot | 33 | ANN | 5 seconds | SBP: 3.23 ± 4.47<br>DBP: 2.73 ± 3.61 |

| Reference | Subjects (F/M) | Wearable device | Biosignals | Physical states | BP Reference | # of BP observations | Model | Window size | MAE ± STD (mmHg) |
|---|---|---|---|---|---|---|---|---|---|
| Liu et al. 2023 [32] | 3077 (2005/1072) | Huawei prototype watch | ECG, PPW, MWPPG | Sitting | Snapshot | 29568 | dMK-BH | 5 seconds | SBP: 9.67 ± 13.04<br>DBP: 6.80 ± 8.78 |
| Wang et al. 2023 [67] | 18 (5/13) | Custom wristband | PPG | Sitting | Snapshot | 309 | MLP | 10 seconds | SBP: — ± 6.00<br>DBP: — ± 6.20 |
| Li et al. 2024 [30] | Training dataset: 438 (192/246)<br><br>Testing dataset: 347 (237/110) | OPPO Watch 3 Pro | PPG | Sitting, Squatting | Training dataset: Snapshot<br><br>Testing dataset: Continuous | — | LR, SVR, RF, LightGBM | 10 or 60 seconds | <u>10 seconds</u><br>SBP: 5.64 ± 7.96<br>DBP: 4.11 ± 5.85<br><br><u>60 seconds</u><br>SBP: 4.46 ± 6.44<br>DBP: 3.22 ± 4.65 |
| Villeneuve et al. 2024 [66] | 15 (6/9) | Custom system | ECG, PPG | Active, Relaxed | Continuous | 903 | SVR, RF, Adaboost | 20 seconds | SBP: 12.74 ± —<br>DBP: 5.99 ± — |
| Li et al. 2025 [29] | 43 (16/27) | Custom wristband | PPG, PPW | Rest, Cold pressor, Breath holding | Continuous | 73175 | MSTNN | 5 seconds | SBP: 6.19 ± 7.89<br>DBP: 4.78 ± 6.23 |
| Van Vliet et al. 2025 [65] | 150 (61/89) | CardioWatch | PPG | At home | Snapshot | 12600 | SVR | 2 minutes | SBP: 3.84 ± 4.46<br>DBP: 4.08 ± 3.97 |

A '—' indicates that the information was not readily reported. ANN: Artificial Neural Network, BLSTM: Bidirectional Long Short-Term Memory (BLSTM), DNN: Deep Neural Network, FNN: Feedforward Neural Network, HNN: Hybrid Neural Network, LightGBM: Light Gradient Boosting Machine, LR: Linear Regression, MLP: Multi-Layer Perceptron, RF: Random Forest, SVR: Support Vector Regression.

## 2 METHODS

### 2.1 HEART-Watch Dataset

The HEART-Watch dataset is comprised of synchronized wrist-based ECG and PPG recordings from a Google Pixel Watch (2022) from 40 participants with 5 snapshot BP readings for each session, resulting in 200 independent observations [25]. These measures were conducted across a semi-naturalistic protocol between sitting, standing, and walking states with realistic wearing conditions and emphasis on a diverse participant cohort. This dataset meets the participant and sex criteria outlined in the revised European Society of Hypertension International Protocol (ESH-IP2), which states that validation datasets should have at least 10 males and females in a minimum cohort size of 33 [41]. However, it should be noted that this dataset was collected on self-reported healthy participants and does not meet the ESH-IP2 minimum cohort size for medium or high BP ranges. The dataset characteristics are listed in Table 2 and a visualization of the BP distribution is provided in Fig 1.

Table 2. HEART-Watch dataset characteristics.

| **Characteristic** | **HEART-Watch Dataset** |
|---|---|
| Participants, n | 40 |
| Sex, male/female | 17/23 |
| Age, years | 44.15 ± 20.81<br>(19-75) |
| Height, meters | 1.69 ± 0.11<br>(1.53 – 1.94) |
| Weight, kilograms | 68.98 ± 5.56<br>(43 – 115) |
| BMI, kg/$m^2$ | 23.85 ± 4.28<br>(16.8 – 35.5) |
| Systolic blood pressure (SBP), mmHg | 122.81 ± 18.22<br>(90 – 182) |
| Diastolic blood pressure (DBP), mmHg | 83.54 ± 11.01<br>(61 – 108) |
| Heart rate (HR), beats/min | 78.71 ± 14.57<br>(50 – 115) |

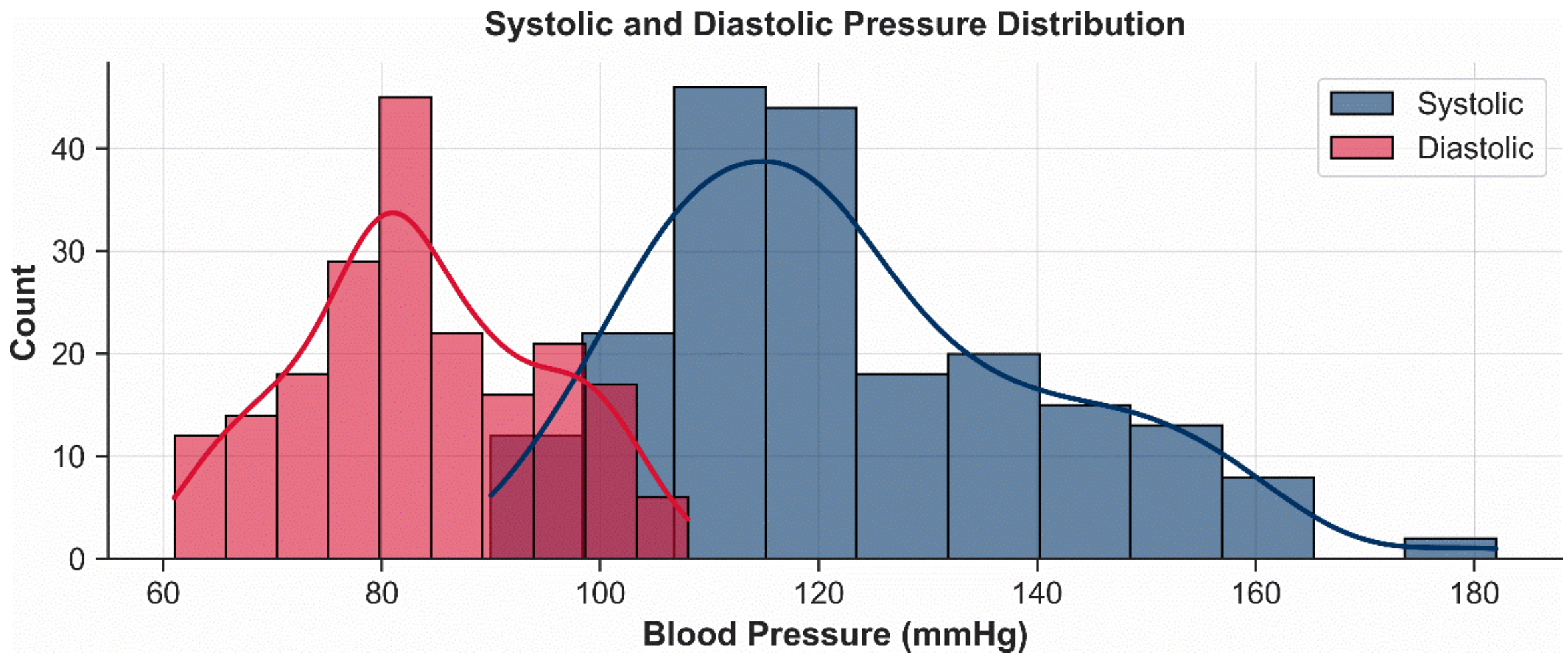

Fig 1. Histogram of SBP and DBP readings for the HEART-Watch dataset with overlaid kernel density estimates.

The original protocol involved variations in physical states (sitting, standing, and walking) that can significantly influence changes in SBP and DBP. To assess these changes, Wilcoxon signed-rank tests were performed to determine whether BP values significantly changed across sequential protocol changes (see Fig 2). After applying a Bonferroni correction for multiple comparisons, it was found that SBP and DBP significantly decreased after 4 minutes of seated rest (Measure 1 → Measure 2), significantly increased when transitioning from sitting to standing (Measure 2 → Measure 3), and significantly increased after 4 minutes of walking (Measure 4 → Measure 5). The statistically significant variations in both SBP and DBP are beneficial for modelling techniques that aim to learn subject-specific BP patterns.

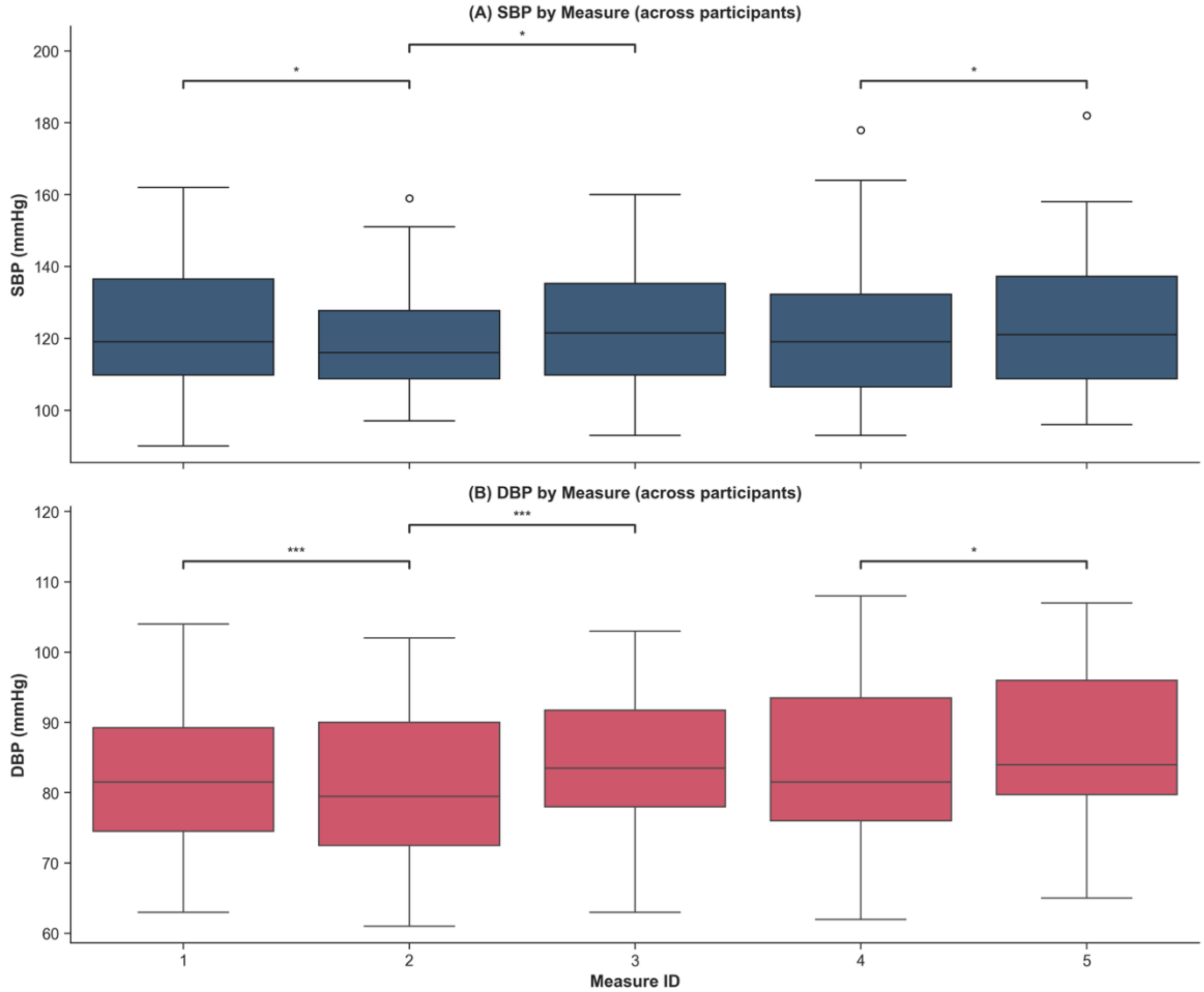

Fig 2. Boxplots and Wilcoxon signed-rank test results of systolic (A) and diastolic (B) blood pressure across measures. * = $p < 0.05$, ** = $p < 0.01$, *** = $p < 0.001$.

## 2.2 Smartwatch ECG

The HEART-Watch dataset provides Lead I smartwatch ECG recordings acquired while participants placed their right index finger on the crown of the watch during BP measurements. Although concurrent chest ECG was also acquired during these sessions, this work focuses on smartwatch ECG to better reflect real-world wearable monitoring situations where only the smartwatch is available. Bland-Altman analysis was conducted to evaluate the agreement between smartwatch ECG and chest ECG morphology through QRS interval measurements. The results of this analysis are provided in Fig 3. Across the five measurement periods, small mean differences were observed between smartwatch ECG and chest ECG QRS intervals, ranging from 0.56 to 1.55 ms. Although the QRS bias from the first measurement period showed statistical significance ($p < 0.05$), the magnitude of this bias was small relative to typical QRS intervals of 80-100 ms. Given the importance of distinguishable QRS complexes for cardiovascular analysis, and the consistently small mean biases across all measurement periods, the smartwatch ECG demonstrated sufficient agreement with chest ECG for feature extraction tasks such as R-peak detection required in this study.

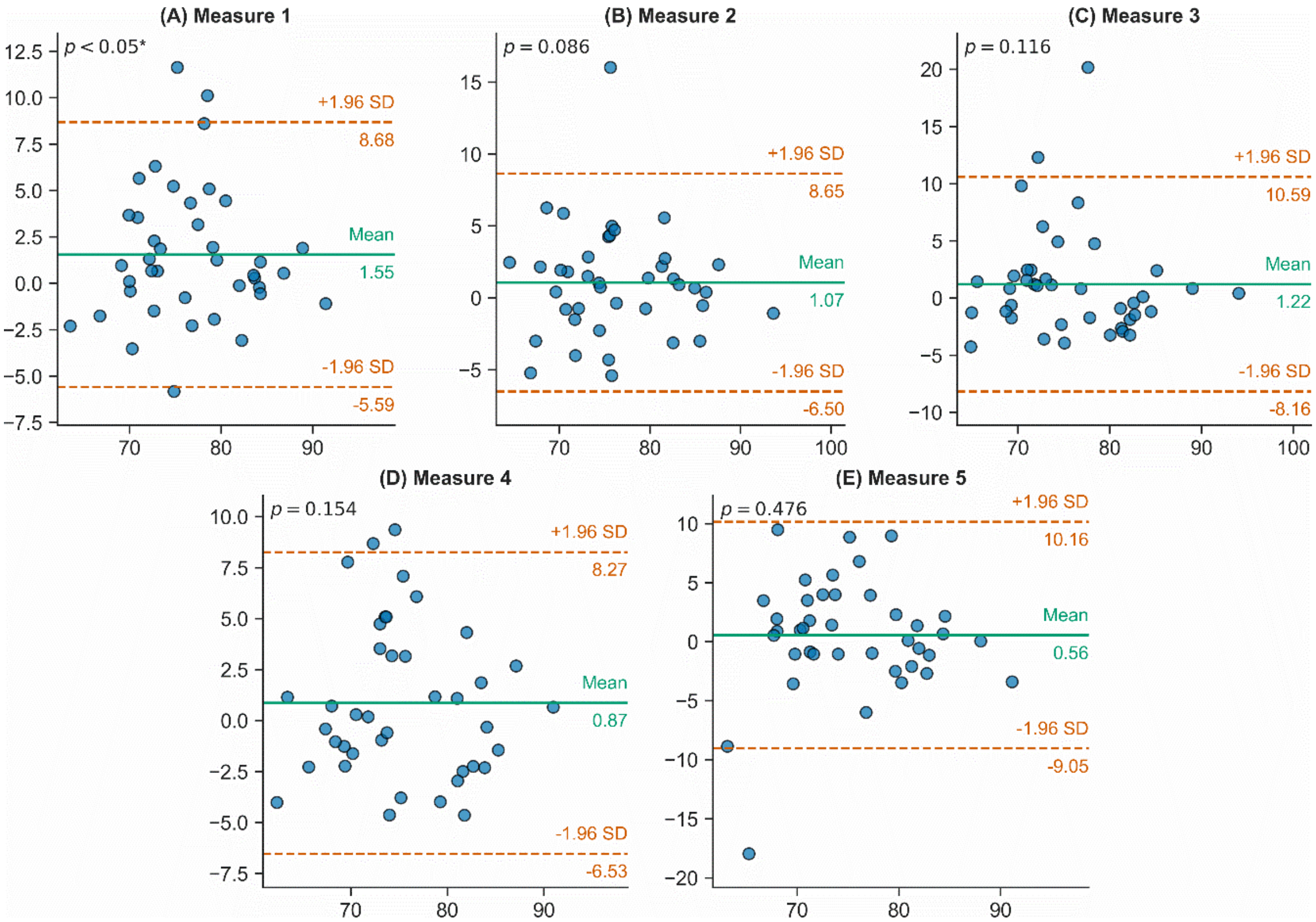

Fig 3. Bland-Altman plots for QRS intervals between smartwatch and chest ECG across all five measurement periods. All measurements are reported in milliseconds (ms).

### 2.3 Signal Preprocessing

The ECG signals were preprocessed using a median filter to remove baseline wander. Baseline wander occurs between 0.8 and 1.25 Hz and represents low-frequency drift that can occur due to electrode shifts, breathing, and other movements [46]. A non-causal, zero-phase, 375$^{th}$ order FIR bandpass filter was applied to ECG signals using a Hamming window between 0.67 and 45 Hz to preserve temporal alignment and QRS complex energy as implemented in the BioSPPy package [5]. Invalid watch ECG recordings were identified by calculating the percentage of recording where window peak-to-peak values were zero. Here, ECG recordings were segmented into 2-second windows and checked for saturation. If more than 10% of a recording contained saturation artifacts, it was considered invalid and substituted with the chest ECG recording. Only subject p03, measure 5 and subject p20, measure 1 were identified as completely invalid recordings and substituted as shown in Fig 4.

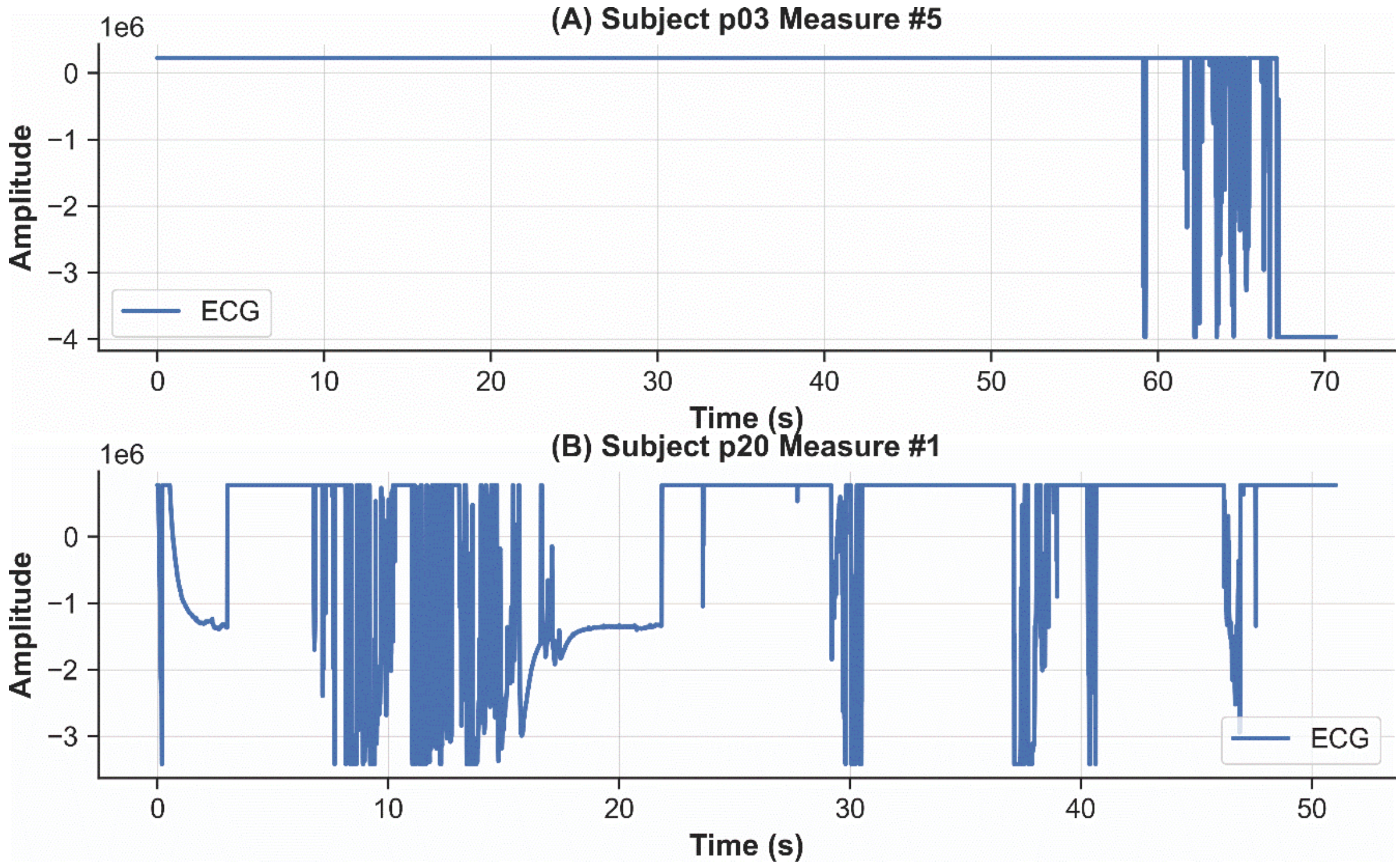

Fig 4. Watch ECG recordings that were marked as invalid due to signal saturation. The square wave morphology is present for the majority of both recordings, indicating that lead fall-off occurred due to improper contact with the electrode. These recordings were substituted with chest ECG recordings.

PPG signals were first inverted because the smartwatch hardware's digital output corresponded to the light input received by the photodiode rather than the pressure-like morphology expected in PPG waveforms [45]. Then, PPG signals were preprocessed with the same median filter to remove baseline wander, followed by bandpass filtering with a zero-phase, 2$^{nd}$ order Butterworth filter between 0.5 and 8 Hz as implemented for peak detection by Elgendi et al. [12]. Both ECG and PPG signals were trimmed by 2 seconds from the start to remove initial recording artifacts and then normalized using Z-score normalization with a sliding window size of 5 seconds with 50% overlap to preserve the waveform shapes. Min-max normalization was not chosen for this process to avoid the influence of sharp amplitude spikes from recording artifacts such as potential lead-fall offs.

### 2.4 Feature Extraction

The synchronized signals were then segmented into 8, 10, 15, 20, 25, and 30 second nonoverlapping windows for feature extraction due to the non-stationary nature of PPG signals [44]. Each window was assigned the same BP label from the entire recording.

#### *2.4.1 Demographic Features*

The demographic for each participant is provided in the HEART-Watch dataset. Specifically, each participant's age, sex, race, height, weight and BMI are numerical and categorical features that can inform predictive model parameters. In addition to numerical BMIs, the Canadian BMI Nomogram was used to generate categorical labels for each participant: Underweight, Normal, Overweight, Obese Class I, and Obese Class II. Boxplots of SBP and DBP per BMI nomogram and sex are shown in Fig 5.

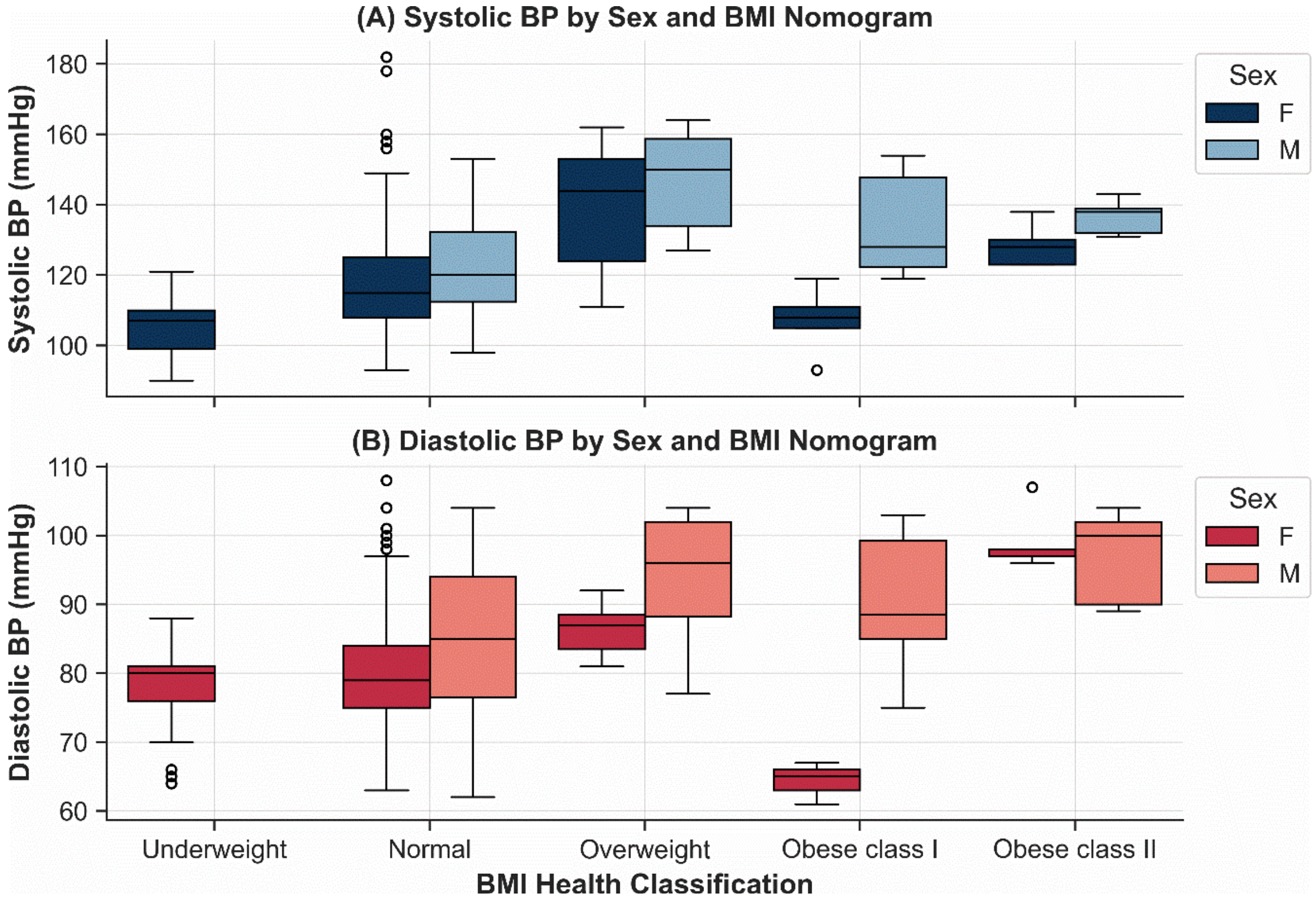

Fig 5. Boxplots of systolic (A) and diastolic (B) blood pressure per BMI classification and sex. F = Female, M = Male.

*2.4.2 Peak and Onset Detection*

Both ECG and PPG waveforms require peak detection algorithms to reliably distinguish local maxima from true R-peak and systolic peaks. For ECG waveforms, this is typically easier given that the R wave has a significantly larger amplitude due to ventricular depolarization compared to the rest of the PQRST complexes. Here, the Peak Prominence ECG Delineator algorithm developed by Emrich et al. is used to perform ECG delineation based on local maxima prominence and physiological constraints [13]. PPG waveforms present more difficulty due to their susceptibility to motion artifacts, ambient lighting, and sensor contact. To address this, the multi-scale peak & trough detection (MSPTD) algorithm was applied to the PPG waveforms to identify true systolic peaks and onsets using scalograms that exclude candidate extrema that don't outperform neighboring candidates [10]. Diastolic features were not examined as wrist-based PPG waveforms significantly vary depending on wrist posture and contact pressure which can reduce distinguishability of signal details such as secondary inflections [22].

*2.4.3 PPG Peak and Onset Selection*

For PPG waveforms, the MSPTD algorithm returned candidate indices for peaks and onsets within each window. These candidate PPG peaks and onsets were then passed through a greedy alignment algorithm. Each onset was paired with the peak that falls within physiologically reasonable limits for crest time, defined as the time interval between the two points. Based on previous and current observations, the minimum crest time was set to 70 ms and the maximum to 500 ms [2, 18]. An example of candidate peaks and onsets is shown in Fig 6, where the candidate points were rejected based on the physiological bounds.

If a candidate onset had multiple candidate peaks prior to the next onset, rather than taking the immediate next peak, the peak with the maximum amplitude was chosen as the corresponding peak for said onset. This logic was chosen to reduce the influence of local maxima that were not the true systolic peaks for a given pulse cycle. Finally, if a candidate onset had another candidate onset prior to the next peak, which can occur due to the presence of multiple local minima, the onset with the lowest amplitude was taken as the onset to be evaluated along with the candidate peak.

Points of maximum slope were then identified by computing the first derivative of each window's PPG segment and extracting the index between each onset and peak with the largest value. The values of each max slope were stored as rates of change per second.

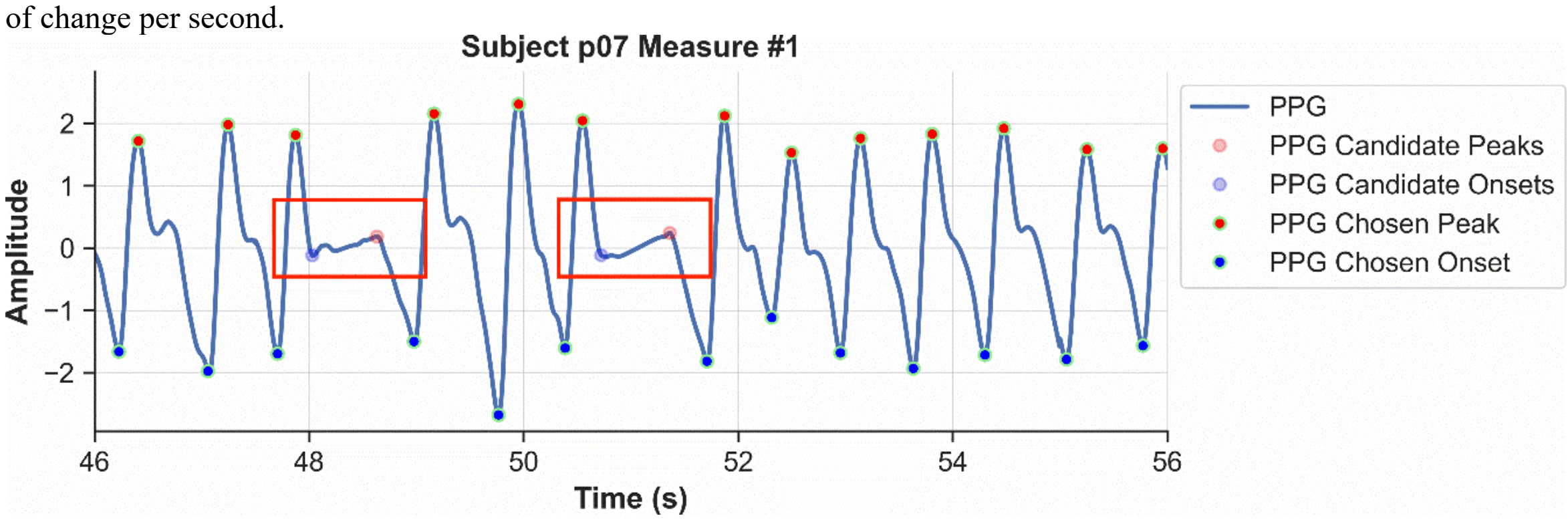

Fig 6. Example of peak and onset selection following initial detection. The red box highlights candidate peaks and onsets that were rejected because their crest times fell outside the specified bounds.

*2.4.4 Pulse Wave Velocity*

The two primary features for cuffless BP estimation through wearable devices are pulse transit time (PTT) and pulse arrival time (PAT). PTT is commonly measured by the time delay of a waveform travelling between two points on the same arterial branch [56]. PAT is computed as the time delays between the R-peaks of the wrist-ECG waveforms and characteristic points of the PPG waveform such as the onsets, maximum slopes, and peaks [56]. The pre-ejection period (PEP) is included in these PAT measurements, and is sensitive to differences in positions, age, and stress [56]. For this work, PAT-based estimation was used due to the availability of synchronized wrist-based ECGs which have been shown to improve PPG-based techniques [47]. Additionally, previous literature has shown that PAT-based estimation may be more robust to changes in physical stress [70]. These PAT measurements are used as the basis for pulse wave velocity theory, where a decrease in PAT is associated with higher blood pressure due to risk factors such as increased arterial stiffness or plaque build-up [38]. By evaluating the timings between cardiac activity at the heart through ECG signals and activity as a peripheral site through PPG signals, BP can be estimated as a reflection of circulatory system performance.

*2.4.5 Pulse Arrival Times*

For each window, PAT values were calculated based on the time between each ECG R-peak and the closest PPG onset, peak, or maximum slope point. To avoid outliers skewing the average PAT measurement per window, physiological bounds were set based on previous and current observations [59]. Given that PPG onsets are expected to occur close to each ECG R-peak, minimum and maximum bounds for valid $PAT_3$ measures were set to 70 ms and 400 ms respectively [28]. The same bounds were used for $PAT_2$ measures. PPG peaks were expected to occur close to the beginning of the following ECG R-peak, so minimum and maximum thresholds for valid $PAT_1$ measures were raised to 100 ms and 650 ms to account for demographic-based differences in arterial stiffness [59]. Fig 7 illustrates how PATs were calculated given the ECG and PPG waveforms within each window.

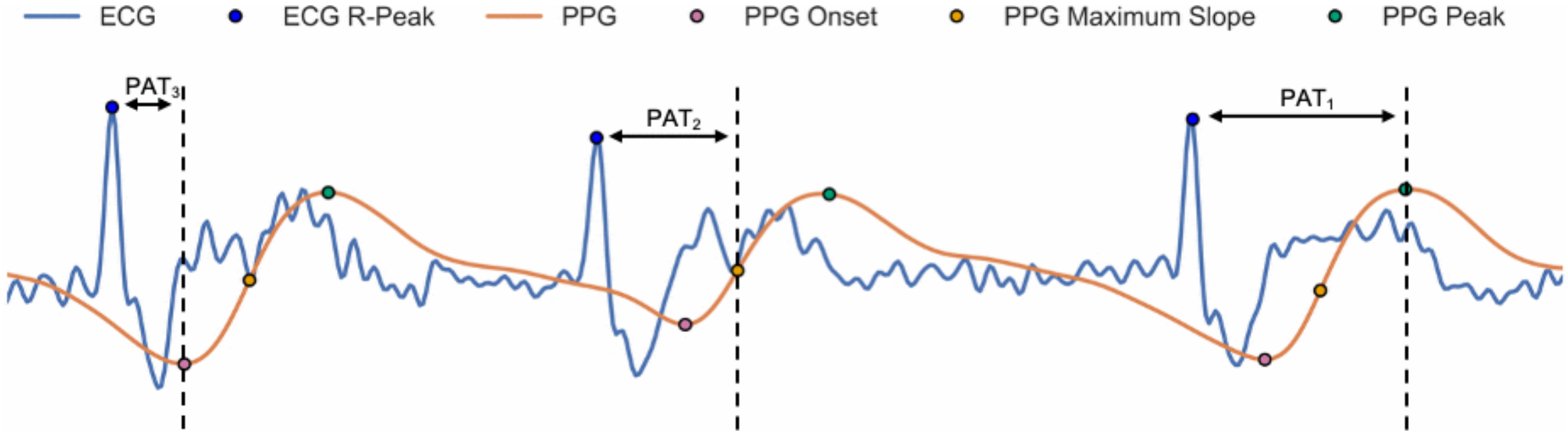

Fig 7 Visualization of how different PAT values are calculated using detected fiducial points in ECG (blue) and PPG (orange): $PAT_1$: PPG cycle peak, $PAT_2$: PPG cycle maximum slope, $PAT_3$: PPG cycle valley.

*2.4.6 ECG Time-domain Features*

ECG waveforms were delineated using the Peak Prominence ECG Delineator algorithm to identify QRS and RR intervals. Specifically, the mean QRS interval for each window's ECG segment was defined as the durations between the onset and offset of each QRS complex. The mean RR interval was calculated using the durations between consecutive R-peaks and, given that the BP cuff reported an HR range of 50-115 BPM, RR intervals were only considered valid between durations corresponding to 35 BPM and 130 BPM. This additional safeguard was used to avoid outlier RR intervals that would likely be incorrectly detected from amplitude spikes caused by recording artifacts.

### 2.5 Calibration-based vs Calibration-free Modelling

Calibration-based modelling includes data from the same subjects in both the training and validation sets [64]. For example, model parameters are calibrated using one or two data points and then validation is performed using the remaining data points for the same subject. Calibration-free modelling, also referred to in this work as generalizable modelling, validates models using features from subjects that were not seen in the training set. Due to the small dataset size and limited number of samples per subject, the focus of this work was calibration-free modelling to highlight the capability of the commercial smartwatch for generalizable BP estimation using unseen subject data [44]. To achieve this, all models were evaluated using leave-one-subject-out cross-validation (LOSO CV), where each subject is treated as the validation set once and the remaining subjects are used for training. The validation technique is ideal for evaluating the realistic performance of BP estimation algorithms on unseen subjects compared to methods such as leave-subjects-out hold out validation where validation splits can significantly alter accuracy [20].

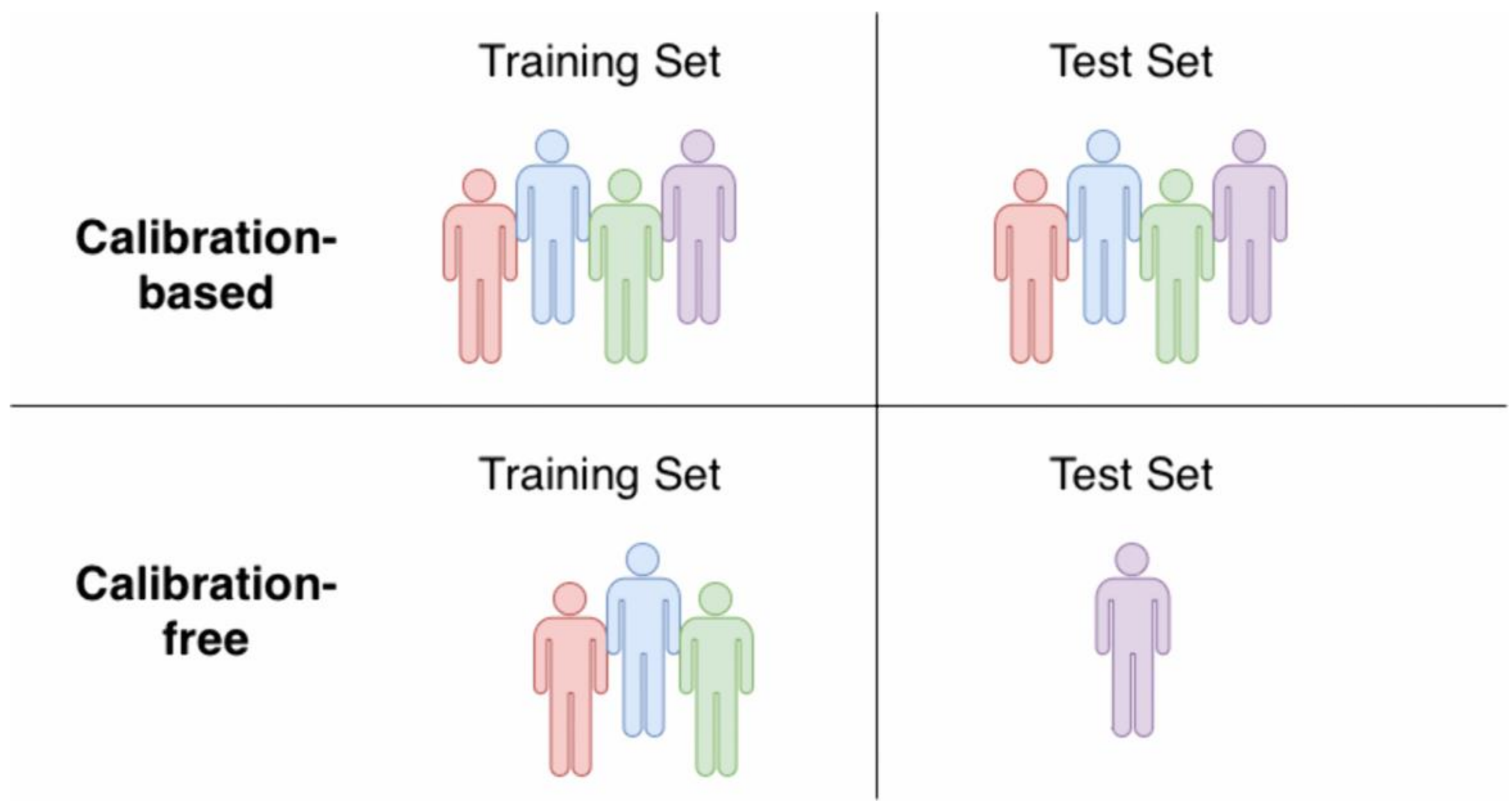

Fig 8. Calibration-based models include features from individuals in both the training and validation sets, while calibration-free or generalizable models only validate on subjects that were excluded from training sets. To achieve calibration-free modelling, LOSO CV was employed in this work.

### 2.6 Mechanism-based Models

Mathematical relationships have been previously used to estimate the relationship between BP and PAT, and are based on vascular elasticity (VE) and elastic tube (ET) models [57]. The VE models incorporate the Moens-Korteweg (MK), Bramwell-Hill (BH), and Hughes equations to relate PAT to mechanical properties of vasculature and blood such as the elastic modulus, radius, area, thickness, and density [32, 57, 72]. The ET modelling approach employs conservation of mass and momentum equations to nonlinearly relate BP to PAT using calibration-specific parameters as well [15, 58].

The mechanism-based models presented in Table 3 were evaluated through LOSO CV, where windows were split using subject IDs to avoid subject leakage between training and validation sets. The mean performance was computed for each window size and model type using mean error (ME), mean absolute error (MAE), and standard deviation (STD) of the signed errors. For models incorporating HR, ECG-derived HR was used rather than BP cuff readings.

Table 3. Existing mathematical models for BP estimation using PAT, where $a$, $b$, $c$, $d$, and $k$ are subject-specific parameters that can be determined through least-squares fitting.

| Mathematical Model | Systolic Blood Pressure (SBP) | Diastolic Blood Pressure (DBP) |
|---|---|---|
| L-MK | $a_s * PAT + b_s$ | $a'_d * PAT + b'_d$ |
| MK-EE | $a_s * \ln(PAT) + b_s$ | $a'_d * \ln(PAT) + b'_d$ |

| R-MK | $a_s * \frac{1}{PAT} + b_s$ | $a'_d * \frac{1}{PAT} + b'_d$ |
|---|---|---|
| HR-MK | $a_s * PAT + b_s * HR_s + c_s$ | $a'_d * PAT + b'_d * HR_s + c'_d$ |
| HR-A-MK | $a_s * PAT + b_s * HR_s + c_s * AGE + d_s$ | $a'_d * PAT + b'_d * HR_s + c'_d * AGE + d'_d$ |
| HR-BMI-MK | $a_s * PAT + b_s * HR_s + c_s * BMI + d_s$ | $a'_d * PAT + b'_d * HR_s + c'_d * BMI + d'_d$ |
| HR-A-BMI-MK | $a_s * PAT + b_s * HR_s + c_s * AGE + d_s * BMI + k_s$ | $a'_d * PAT + b'_d * HR_s + c'_d * AGE + d'_d * BMI + k'_d$ |
| M-M | $a_s + \sqrt{b_s + c_s * \frac{1}{PAT^2}}$ | $a_d + \sqrt{b_d + c_d * \frac{1}{PAT^2}}$ |

## 2.7 Machine Learning Models

One of the challenges of implementing a generalizable cuffless BP estimation model using mechanism-driven approaches is that underlying assumptions about the mechanical properties of human peripheral vasculature are too constrained given the dynamic changes that occur during cardiac cycles [31, 32, 36]. Additionally, these models may not be complex enough to generalize to diverse cohort sizes given their limited number of calibration parameters and the number of BP labels available in this dataset specifically. Traditional machine learning (ML) methods can offer more flexible modelling techniques that incorporate handcrafted features into model estimations [44]. The 15 handcrafted features derived in the Feature Extraction section of this work are summarized in Table 4. These features are intended to capture waveform morphology and demographic characteristics associated with cardiovascular health, including blood viscosity and peripheral vascular resistance. For example, PPG-derived features, such as crest time and amplitude of maximum slope, characterize PPG morphology on a beat-by-beat basis which can provide information related to individual vascular compliance [44]. Although some features are correlated, they are commonly used for cuffless BP monitoring applications because they capture complementary details about cardiovascular physiology [72]. Demographic information such as height and weight provides quantitative descriptors of body size, while derived BMI provides a descriptor of obesity that can inform BP estimations [23].

The Regression Learner App in MATLAB's Statistics and Machine Learning Toolbox was used to evaluate 17 supervised ML models for systolic and diastolic estimation across all window sizes. Models were trained and validated using LOSO CV. Performance metrics were measured through ME, MAE, and STD to evaluate how each ML model performed on unseen data.

### *2.7.1 Kernel-based Models*

Kernel-based regression models were chosen to capture nonlinear relationships between BP and input features that linear regression models may not represent. Kernel-based models utilize the "kernel trick", which is a technique that maps input features from a low-dimensional space to a high-dimensional space, and fits a linear model. This high-dimensional linear model corresponds to a nonlinear solution in the original low-dimensional space.

Gaussian process regression (GPR) models, support vector machines (SVMs), and kernel approximation models were evaluated. GPR models provide probability distributions of potential functions to fit the data, with each kernel representing the family of functions for modelling. SVMs, or support vector regression (SVR), utilize kernels functions to model nonlinear relationships by modeling a linear function in the higher-dimension space. In kernel approximation models, rather than using an exact kernel function like SVMs, the nonlinear mapping is approximated by generating nonlinear features and fitting a linear model.

*2.7.2 Tree-based models*

Regression trees are nonlinear models that offer strong interpretability due to their decision tree format but can be prone to overfitting, especially on small datasets. Coarse, medium and fine single regression trees were evaluated, where the descriptor refers to the minimum number of training samples present in terminal/leaf nodes. Fine trees with smaller leaf sizes tend to show lower bias and higher variance, whereas coarse trees with larger leaf sizes exhibit higher bias and lower variance.

In addition to single trees, ensembles of trees were evaluated using boosted and bagged trees. Boosted trees were trained using least-squares boosting (LSBoost), an adaptive boosting (AdaBoost) algorithm for regression ensembles that sequentially builds a strong predictive model by fitting new trees to the residuals from previous, weaker models. Bagged trees employ bootstrapping and aggregating (bagging), where multiple trees are trained using subsets of the input and averaged to yield a single model output.

Table 4. Handcrafted features derived from ECG and PPG waveforms.

| **Feature Group** | **Biosignals** | **Feature** |
|---|---|---|
| Demographic | N/A | Age (years) |
| | | Sex (M/F) |
| | | Race (categorical) |
| | | Height (cm) |
| | | Weight (kg) |
| | | BMI (kg/m$^2$) |
| | | BMI Class (categorical) |
| Time-domain | ECG | RR Interval (s) |
| | | QRS Interval (s) |
| | | Heart Rate (beats/min) |
| | PPG | Crest time (s) |
| | | Amplitude of maximum slope |
| | ECG, PPG | $PAT_1$: Average ECG R-peak to PPG peak (s) |
| | | $PAT_2$: Average ECG R-peak to PPG maximum slope (s) |
| | | $PAT_3$: Average ECG R-peak to PPG onset (s) |

### 2.8 Deep Learning Models

While traditional ML methods address some of the shortcomings of mechanism-driven modelling, they require complex feature extraction pipelines in order to derive suitable inputs to models [4, 14, 62]. For wrist-based signal acquisition across a diverse range of participants, complex feature engineering may not be suitable given the quality concerns of acquired biosignals. For example, PPG waveforms can exhibit significant variations in morphology depending on sensor contact, skin elasticity, and skin pigment [44]. As a result, the expectations of feature extraction pipelines for ideal signal morphologies may not be the reality during actual measurements with commercial smartwatches. Even when signal quality is ideal, the necessary combination of features for optimal feature sets may not always be known without laborious experimentation. To examine if performance can be improved beyond mechanism-driven and ML modelling, a deep learning (DL) model was developed to accept wrist-based ECG and PPG time-series, in addition to demographic information, and predict SBP and DBP.

#### *2.8.1 Model Architecture*

The proposed DL architecture consists of four primary components: a signal encoder, demographics encoder, demographic-wise linear signal modulation layer, and a prediction layer.

The signal encoder was comprised of two dilated convolutional neural network (CNN) blocks separated by a strided convolution layer. Dilated convolution was chosen based on the architecture proposed by Baek et al., where parallel dilated convolutions with concatenated outputs were shown to enhance temporal feature extraction through expanded receptive fields [4]. In our work, the first dilated convolution operated independently on ECG and PPG signals to extract modality-specific features using four outputs channels per modality. The resulting branches were then concatenated and fused using a 1D pointwise convolution with a kernel size of 1, followed by GroupNorm and ReLU activation. The fused features were subsequently downsampled by 50% using a strided convolution (stride of 2), as proposed by Baek et al. to reduce information loss associated with traditional pooling operations [4]. The downsampled features were then passed to the second dilated convolution block, where cross-modality relationships between the ECG and PPG inputs are learned. Dilated convolutions used dilation factors of 1, 2, 4, and 6.

The demographics encoder was a lightweight MLP composed of two fully connected layers. Demographic variables were projected through the first linear layer to 32 hidden units, followed by ReLU activation and 20% Dropout for regularization. The second linear layer then mapped these representations to a 64-feature embedding with ReLU activation.

The demographic-wise linear signal modulation layer was inspired by previous work from Perez et al., which introduced feature-wise linear modulation (FiLM) to adaptively modulate intermediate features using learned affine transformations [49]. In our work, a FiLM layer was applied to linearly scale and shift the encoded signal features as a function of demographic embeddings. This workflow is similar to how FiLM techniques have previously been used for birthweight prediction tasks [39]. As recommended by Perez et al., a single linear layer was used as the FiLM generator, mapping demographic embeddings to scale and shift parameters. To prevent unbounded feature scaling, a hyperbolic tangent (Tanh) activation function was applied to scaling parameters prior to feature modulation. The motivation behind this design was to use demographic information as a conditioning signal for adjusting signal representations. Demographic factors such as age have been shown to influence PPG morphology [2, 11], and population-level trends such as those shown in Fig 5 suggest that baseline BP values can vary significantly between sub-groups.

Finally, the modulated signal encodings were pooled to the match the window length (e.g., a 15-second window yields 15 features) and flattened before being passed to the prediction layer. The prediction layer was an MLP that applied

LayerNorm, followed by a linear layer with 64 hidden neurons and ReLU activation. A second linear layer then maps these representations to 2 outputs corresponding to SBP and DBP predictions. The full architecture is visualized in Fig 9.

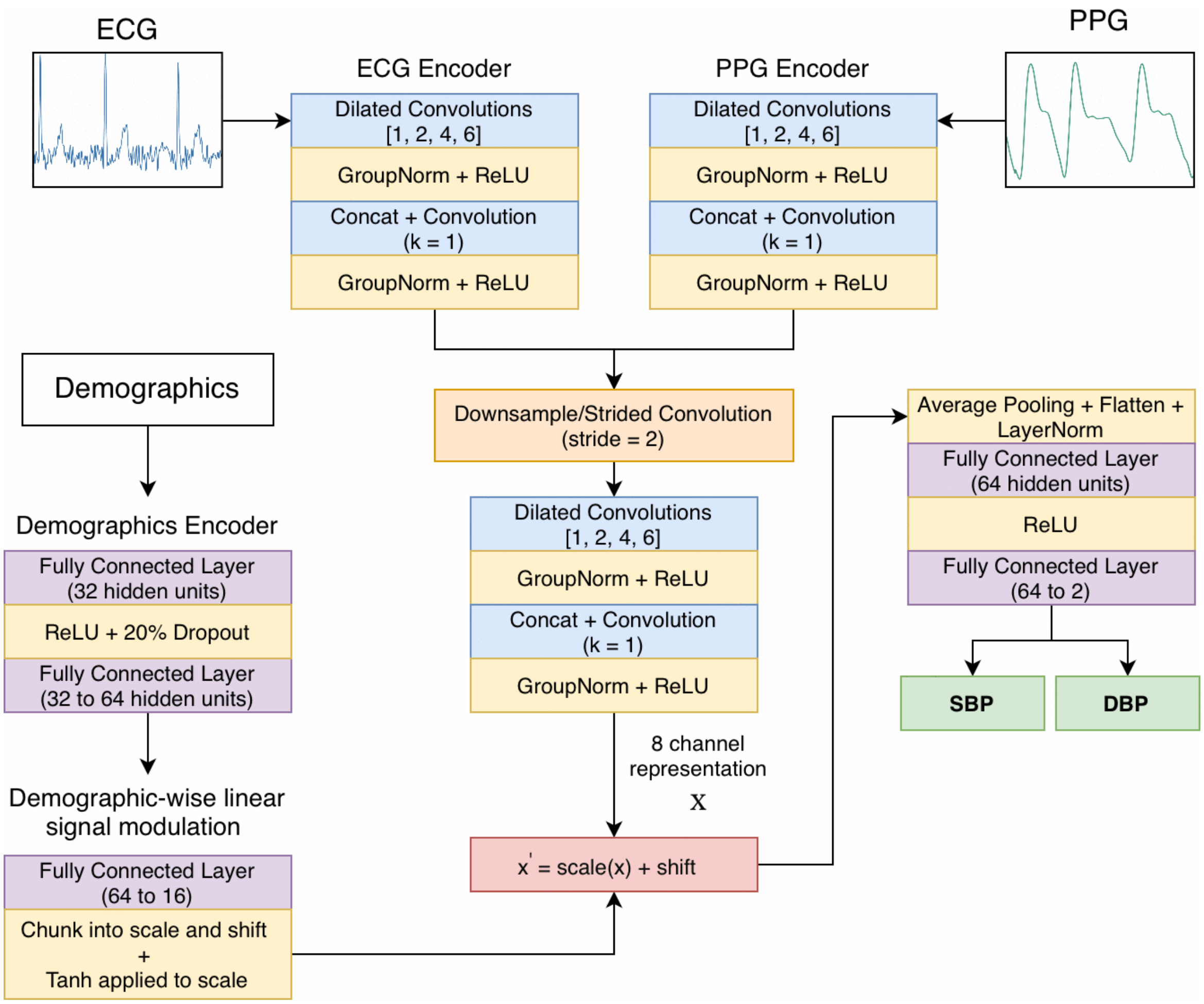

Fig 9. Visualization of proposed DL architecture to predict SBP and DBP using ECG and PPG time-series windows and demographic information.

*2.8.2 Data Preparation*

The categorical demographic inputs (sex, race, and BMI class) were one-hot encoded to prevent the model from inferring ordinal relationships between categories. Continuous demographic variables (age, height and weight) were normalized using statistics computed from training splits, and the same normalization parameters were applied to both training and test data to avoid data leakage. Numerical BMI values were excluded from the demographic feature set based on preliminary experimentation. ECG and PPG signals were preprocessed as described in the Signal Preprocessing section and segmented into non-overlapping windows of 8, 10, 15, 20, 25, and 30 seconds as before.

### *2.8.3 Training Settings*

Train and test splits were generated using LOSO CV, where windows from the held-out subject were reserved for testing and the remaining 39 subjects were used for training. The AdamW optimizer was used with a learning rate of 3 x $10^{-3}$ and weight decay of 1 x $10^{-3}$. L1 loss was selected as the training objective to directly minimize MAE, while a scheduler reduced the learning rate by 90% if the training loss did not improve after 5 epochs. Gradients were clipped to a maximum norm of 1.0 following backpropagation and prior to weight updates to improve training stability. Each fold was trained for 20 epochs, and the model checkpoint with the lowest combined SBP and DBP MAE on the training set was selected for evaluation on the held-out subject.

## 2.9 Model Evaluation

### *2.9.1 Baseline and Naïve Comparison*

The best performing model was compared against both a baseline and a naïve estimator. For each LOSO fold, the baseline estimator predicted constant value SBP and DBP values corresponding to the mean BP of the training population. The naïve estimator predicted fixed SBP and DBP values of 120 and 80 mmHg, respectively, representing a clinically motivated normal BP value based on established guidelines [68]. Performance was evaluated using the same metrics as the best performing model (MAE, ME, and STD). These comparisons were conducted to assess whether the proposed model learned meaningful demographic- and biosignal-specific relationships for BP estimation rather than collapsing towards population-level means [60].

### *2.9.2 International Standards*

The best performing model from the investigation was evaluated using the standards established by the Association for the Advancement of the Medical Instrumentation (AAMI) which requires BP estimation devices to achieve ME and STD within ± 5 mmHg and ± 8 mmHg respectively [69]. Additionally, the MAE values were used to evaluate candidate models using the IEEE 1708 standard, where Grades A to D are assigned based on MAE performance [74]. Similarly, grading criteria published by the British Hypertension Society (BHS) were used to evaluate cumulative MAE performance [42], where all cumulative percentages must be greater than or equal to the thresholds associated with each grade. Here, a device is recommended if both SBP and DBP estimation passes the AAMI criteria and receives grade A or B using the BHS protocol. A summary of the standards used for evaluating cuffless BP estimation algorithms are shown in Table 5.

Table 5. Summary of AAMI, IEEE 1708, and BHS standards for benchmarking BP estimation devices.

| | **AAMI** | **IEEE 1708** | **BHS** | | |
|---|---|---|---|---|---|
| **Recommended Grade** | **ME ± STD** | **MAE** | **MAE ≤ 5 mmHg (%)** | **MAE ≤ 10 mmHg (%)** | **MAE ≤ 15 mmHg (%)** |
| Grade A/Pass | 5 ± 8 mmHg | ≤ 5 mmHg | 60 | 85 | 95 |
| Grade B | — | ≤ 6 mmHg | 50 | 75 | 90 |
| Grade C | — | ≤ 7 mmHg | 40 | 65 | 85 |
| Grade D | — | ≥ 7 mmHg | Worse than Grade C | | |

*2.9.3 Ablation Study*

To assess the contribution of each input modality to model performance, ablation analyses were conducted using the best model configuration. Input modalities were systematically removed, and the model was retrained under identical LOSO CV conditions. In addition to the primary performance metrics of ME, MAE, and STD, each ablated model variant was further evaluated using the AAMI, IEEE 1708, and BHS standards.

*2.9.4 Subgroup Study*

A subgroup analysis was conducted to evaluate the model performance across different BMI subgroups. Participants were classified into obese and non-obese groups using a BMI threshold of 30 $kg/m^2$, where the obese group included Obese Class I and II individuals, while the non-obese group included Underweight, Normal, and Overweight individuals. The best model configuration was evaluated using the same LOSO CV framework using the primary performance metrics of ME, MAE, and STD. Statistical significance between subgroups was assessed using a two-sided Mann-Whitney U test due to the small sample size and non-Gaussian error distributions. This non-parametric test compares pairwise observations to determine whether MAE values consistently differ between participants in each subgroup without assuming directionality. Effect size was quantified using Cliff's δ as a non-parametric measure of subgroup differences.

# 3 RESULTS

## 3.1 Correlation between Blood Pressure and Pulse Arrival Time

The computed PATs were statistically evaluated against their corresponding BP labels using Pearson correlation coefficients to assess the relationship between smartwatch-derived PATs and SBP/DBP. The results are visualized in Fig 10. Previous studies have generally reported stronger correlations between PAT and SBP than with DBP when derived from ECG and PPG signals [73].

For $PAT_1$, calculated from the ECG R-peak to the PPG systolic peak, a statistically significant positive correlation was observed with SBP ($SBP_{PAT_1}$: $r = 0.30$, $p = 1.5 \times 10^{-5}$), while the relationship with DBP was weaker and not statistically significance ($DBP_{PAT_1}$: $r = 0.13$, $p = 0.066$). Similarly, $PAT_3$, derived from the ECG R-peak to the PPG onset, exhibited statistically significance negative correlations with both SBP ($SBP_{PAT_3}$: $r = -0.30$, $p = 1.5 \times 10^{-5}$) and DBP ($DBP_{PAT_3}$: $r = -0.22$, $p = 0.0014$). These findings are consistent with the expected inverse relationship between BP and PAT, where increasing BP is associated with faster pulse propagation and shorter transit times. In contrast, $PAT_2$, computed using the ECG R-peak and PPG point of maximum slope, did not exhibit statistically significant linear correlations with either SBP ($SBP_{PAT_2}$: $r = 0.04$, $p = 0.6$) or DBP ($DBP_{PAT_2}$: $r = 0.00$, $p = 0.95$). This suggests that PPG onsets and peaks may serve as more valuable fiducial points than maximum slope locations for wrist-based BP estimation.

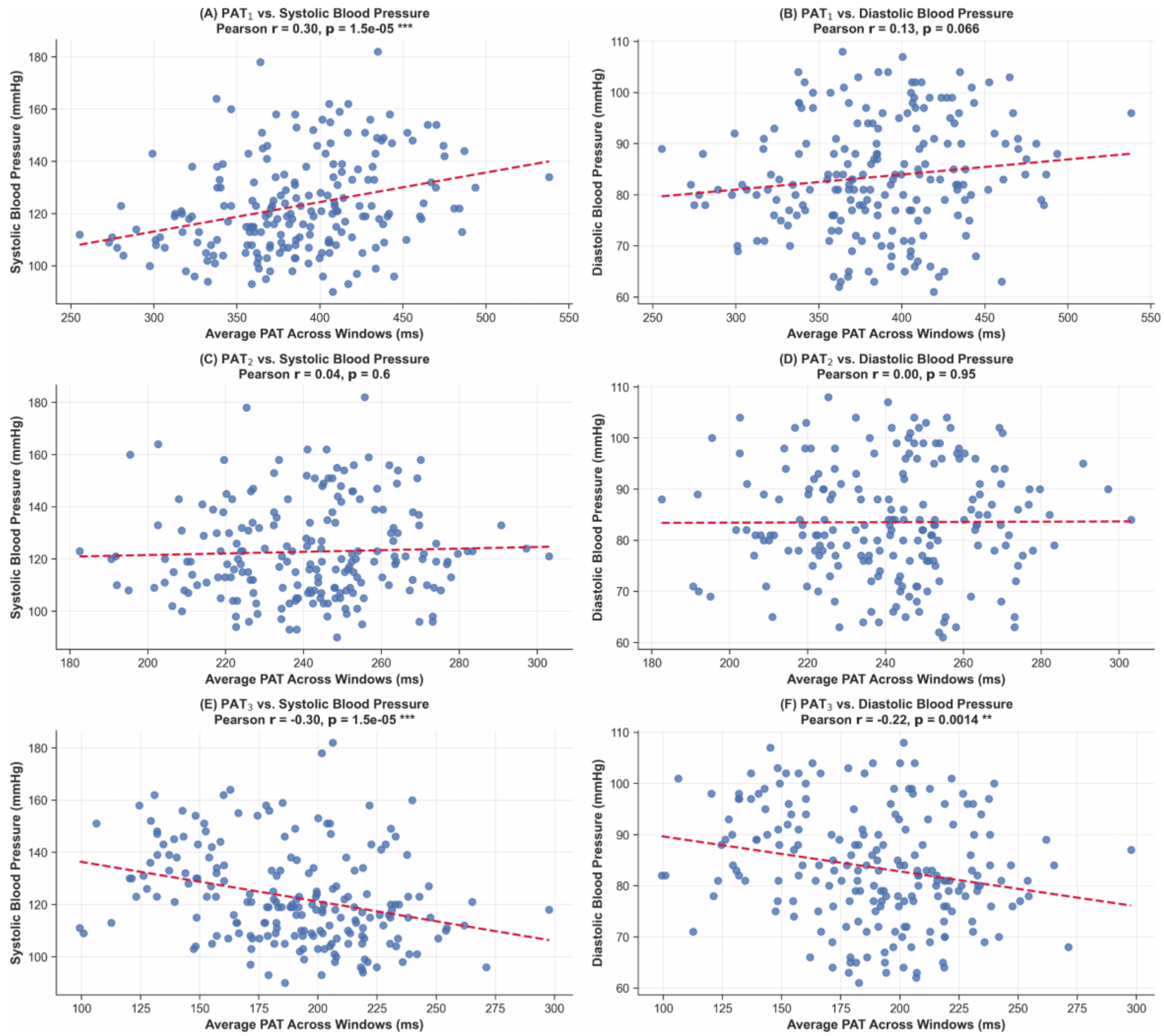

Fig 10. Scatterplot of systolic and diastolic blood pressure by PAT measurements computed across windows with no segmentation. Both $PAT_1$ and $PAT_3$ exhibit significant linear correlations with blood pressure. * = $p < 0.05$, ** = $p < 0.01$, *** = $p < 0.001$.

### 3.2 Mechanism-based Models

The results of LOSO CV for mechanism-based modelling are shown in Table 6. Across all window sizes, models incorporating demographic features consistently outperformed models relying solely on signal-derived features. Specifically, the HR-A-BMI-MK and HR-BMI-MK models achieved the lowest errors for SBP and DBP estimation, respectively. For SBP estimation, the time interval between the ECG R-peak and PPG point of maximum slope ($PAT_2$) consistently yielded the lowest MAE across all window sizes, despite exhibiting weak linear correlations with SBP. For DBP estimation, both $PAT_2$ and $PAT_3$ achieved comparable performance with similar MAE values across all window lengths. Overall, model performance remained relatively stable across window sizes, although a slight reduction in SBP

MAE was observed with increasing window length. Considering both MAE and STD, the 30-second window yields the best mechanism-driven model (SBP: 11.88 ± 6.30, DBP: 8.75 ± 3.98).

Table 6. Results of LOSO CV using mechanism-based models. The best models per window size were defined as those that yielded the lowest average MAE.

| Window Size | Systolic Blood Pressure (SBP) | | | | | Diastolic Blood Pressure (DBP) | | | | |
|---|---|---|---|---|---|---|---|---|---|---|
| | Best model | PAT | MAE (mmHg) | ME (mmHg) | STD (mmHg) | Best model | PAT | MAE (mmHg) | ME (mmHg) | STD (mmHg) |
| 8 seconds | HR-A-BMI-MK | $PAT_2$ | 12.27 | 0.34 | 6.34 | HR-BMI-MK | $PAT_2$ | 8.68 | -0.07 | 4.02 |
| 10 seconds | HR-A-BMI-MK | $PAT_2$ | 12.11 | 0.31 | 6.56 | HR-BMI-MK | $PAT_2$ | 8.68 | -0.06 | 4.10 |
| 15 seconds | HR-A-BMI-MK | $PAT_2$ | 12.15 | 0.47 | 6.42 | HR-BMI-MK | $PAT_3$ | 8.69 | 0.01 | 4.12 |
| 20 seconds | HR-A-BMI-MK | $PAT_2$ | 11.86 | 0.37 | 6.50 | HR-BMI-MK | $PAT_3$ | 8.70 | -0.06 | 4.17 |
| 25 seconds | HR-A-BMI- MK | $PAT_2$ | 11.88 | 0.19 | 6.45 | HR-BMI-MK | $PAT_3$ | 8.68 | -0.17 | 4.16 |
| 30 seconds | HR-A-BMI-MK | $PAT_2$ | 11.88 | 0.44 | 6.30 | HR-BMI-MK | $PAT_2$ | 8.75 | -0.01 | 3.98 |

### 3.3 Machine Learning Models

The results of LOSO CV for ML models are presented in Table 7. ML approaches achieved slightly lower SBP MAE values than mechanism-based models, ranging from 11.08 to 11.23 mmHg. For DBP estimation, MAE values were comparable to or slightly worse than mechanism-based approaches, ranging from 8.78 to 9.13 mmHg. Among the 17 evaluated models, Coarse Gaussian SVM and Quadratic SVM most frequently produced the lowest SBP errors, while Exponential GPR usually performed best for DBP estimation. Despite achieving slightly lower MAE values, all ML models exhibited substantially higher STD values compared to mechanism-based models across all window sizes (SBP STD ≈ 14 mmHg, DBP STD ≈ 11 mmHg). This finding suggests that although ML models reduced average prediction errors, they were more sensitive to inter-subject variability, resulting in inconsistent predictions for held-out subjects. Furthermore, these STD values exceed the AAMI criteria for BP estimation devices (STD ≤ 8 mmHg), indicating that these approaches were unsuitable for calibration-free BP estimation in this cohort.

Table 7. Results of LOSO CV using ML models. The best models per window size were defined as those that yielded the lowest average MAE.

| Window Size | Systolic Blood Pressure (SBP) | | | | Diastolic Blood Pressure (DBP) | | | |
|---|---|---|---|---|---|---|---|---|
| | Best model | MAE (mmHg) | ME (mmHg) | STD (mmHg) | Best model | MAE (mmHg) | ME (mmHg) | STD (mmHg) |
| 8 seconds | Coarse Gaussian SVM | 11.17 | 0.45 | 14.21 | Exponential GPR | 8.79 | 0.83 | 10.79 |

| 10 seconds | Coarse Gaussian SVM | 11.08 | 0.43 | 14.10 | Linear SVM | 8.78 | 0.63 | 10.96 |
|---|---|---|---|---|---|---|---|---|
| 15 seconds | Coarse Gaussian SVM | 11.15 | 0.56 | 14.17 | Exponential GPR | 8.89 | 0.95 | 10.93 |
| 20 seconds | Exponential GPR | 11.19 | 1.68 | 14.15 | Exponential GPR | 8.98 | 1.17 | 10.90 |
| 25 seconds | Quadratic SVM | 11.23 | 0.72 | 14.53 | Coarse Gaussian SVM | 9.02 | 0.66 | 11.16 |
| 30 seconds | Quadratic SVM | 11.23 | 0.66 | 14.57 | Exponential GPR | 9.13 | 0.87 | 11.02 |

### 3.4 Deep Learning Model

The results of LOSO CV using the proposed DL approach are shown in Table 8. The proposed architecture achieved the lowest MAE observed across all three modelling approaches, with the best performance obtained using an 8-second window (SBP MAE: 9.13 mmHg, DBP MAE: 7.77 mmHg). Across window lengths, no consistent improvement was observed with increasing window durations, while SBP errors increase slightly for longer windows. The proposed DL approach also produced the lowest STD values compared to both mechanism-based and ML models, while remaining stable across window lengths (SBP STD ≈ 5.50 mmHg, DBP STD ≈ 3.80 mmHg). This combination of lower variability and reduced MAE suggests that the DL model was more robust to inter-subject variability, reducing the large prediction variability observed in ML approaches while maintaining improved predictive performance over mechanism-based models.

Table 8: Results of LOSO CV using the proposed DL model. The best models per window size were defined as those that yielded the lowest average MAE.

| Window Size | Systolic Blood Pressure (SBP) | | | Diastolic Blood Pressure (DBP) | | |
|---|---|---|---|---|---|---|
| | MAE (mmHg) | ME (mmHg) | STD (mmHg) | MAE (mmHg) | ME (mmHg) | STD (mmHg) |
| 8 seconds | 9.13 | 0.06 | 5.60 | 7.77 | -0.14 | 3.84 |
| 10 seconds | 10.02 | 0.50 | 5.54 | 7.84 | 0.57 | 3.81 |
| 15 seconds | 9.38 | 0.03 | 5.54 | 7.53 | 0.57 | 3.82 |
| 20 seconds | 10.20 | 0.35 | 5.45 | 8.42 | 0.85 | 3.81 |
| 25 seconds | 10.41 | 1.96 | 5.46 | 8.21 | 1.42 | 3.81 |
| 30 seconds | 10.28 | 0.77 | 5.42 | 8.03 | 1.20 | 3.73 |

### 3.5 Baseline and Naïve Comparison Results

The proposed DL model outperformed the baseline and naïve estimator for both SBP and DBP estimation (see Table 9). These improvements were primarily driven by reductions in MAE, while STD values remained comparable between models. These findings suggest that the proposed model did not collapse towards constant predictions and instead leveraged demographic and biosignal inputs to capture individualized BP variation.

Table 9. Results of LOSO CV comparing the best performing DL model, a baseline estimator predicting the training population mean, and a naïve estimator predicting a fixed value of 120/80 mmHg.

| Model | Systolic Blood Pressure (SBP) | | | Diastolic Blood Pressure (DBP) | | |
|---|---|---|---|---|---|---|
| | MAE (mmHg) | ME (mmHg) | STD (mmHg) | MAE (mmHg) | ME (mmHg) | STD (mmHg) |
| DL model (this work) | 9.13 | 0.06 | 5.60 | 7.77 | -0.14 | 3.84 |
| Training Population Mean | 15.14 | 0.67 | 5.48 | 9.28 | 0.10 | 3.80 |
| Naïve Estimator | 14.33 | -2.79 | 5.48 | 9.26 | -3.45 | 3.80 |

### 3.6 Ablation Study Results

Based on the results of all three modelling approaches, the proposed DL model outperformed the other approaches, with the best performance achieved using an 8-second window. Benchmarking results for this model using the AAMI, IEEE 1708, and BHS standards are shown for SBP (Table 10) and DBP (Table 11) estimation, respectively.

Table 10. Results of ablation analyses on the best performing DL model for SBP estimation using AAMI, IEEE, and BHS standards across all participants.

| | AAMI | | IEEE 1708 | | BHS | | | |
|---|---|---|---|---|---|---|---|---|
| Model Inputs | ME ± STD (mmHg) | Pass/Fail | MAE (mmHg) | Resulting Grade | MAE ≤ 5 mmHg (%) | MAE ≤ 10 mmHg (%) | MAE ≤ 15 mmHg (%) | Resulting Grade |
| ECG + PPG + Demographics | 0.06 ± 5.60 | Pass | 9.13 | Grade D | 32.6 | 60.7 | 80.0 | Grade D |
| ECG + Demographics | -0.11 ± 5.57 | Pass | 9.30 | Grade D | 32.1 | 59.2 | 77.7 | Grade D |
| PPG + Demographics | -0.43 ± 5.52 | Pass | 10.90 | Grade D | 25.6 | 52.0 | 73.9 | Grade D |
| ECG + PPG | -0.79 ± 7.78 | Pass | 12.99 | Grade D | 24.1 | 46.6 | 62.6 | Grade D |
| ECG | -0.35 ± 7.07 | Pass | 13.57 | Grade D | 23.0 | 45.0 | 61.0 | Grade D |
| PPG | -1.88 ± 7.48 | Pass | 14.55 | Grade D | 19.6 | 39.9 | 57.9 | Grade D |

Table 11. Results of ablation analyses on the best performing DL model for DBP estimation using AAMI, IEEE, and BHS standards across all participants.

| | **AAMI** | | **IEEE 1708** | | **BHS** | | | |
|---|---|---|---|---|---|---|---|---|
| **Model Inputs** | **ME ± STD (mmHg)** | **Pass/Fail** | **MAE (mmHg)** | **Resulting Grade** | **MAE ≤ 5 mmHg (%)** | **MAE ≤ 10 mmHg (%)** | **MAE ≤ 15 mmHg (%)** | **Resulting Grade** |
| ECG + PPG + Demographics | -0.14 ± 3.84 | Pass | 7.77 | Grade D | 40.0 | 68.7 | 86.9 | Grade C |
| ECG + Demographics | -0.10 ± 3.85 | Pass | 7.54 | Grade D | 39.4 | 68.8 | 85.9 | Grade D |
| PPG + Demographics | -0.39 ± 3.80 | Pass | 8.25 | Grade D | 35.9 | 62.9 | 84.1 | Grade D |
| ECG + PPG | -0.33 ± 5.27 | Pass | 9.96 | Grade D | 27.0 | 54.9 | 76.6 | Grade D |
| ECG | -0.76 ± 4.71 | Pass | 10.63 | Grade D | 25.8 | 49.1 | 71.1 | Grade D |
| PPG | -0.50 ± 5.33 | Pass | 9.37 | Grade D | 32.1 | 55.9 | 77.0 | Grade D |

The ablation study systematically evaluated the contribution of each input modality – ECG, PPG, and demographic features – to the DL model's estimation performance. For SBP estimation, the complete model (ECG + PPG + Demographics) achieved the best overall performance (ME: 0.06 mmHg, MAE: 9.13 mmHg, STD: 5.60 mmHg), passing the AAMI standard while receiving Grade D under both IEEE 1708 and BHS criteria. Removing PPG inputs resulted in only a slight reduction in performance (MAE: 9.30 mmHg), whereas removing ECG inputs produced larger degradations (MAE: 10.90 mmHg). This suggests that ECG waveforms contribute more strongly for SBP generalization than PPG signals. The largest performance degradation occurred when demographic features are removed, increasing MAE from 9.13 to 12.99 mmHg and STD from 5.60 to 7.78 mmHg. Additional ablation analyses further supported this trend, where the PPG-only model produced the worst SBP performance overall (ME: -1.88 mmHg, MAE: 14.55, STD: 7.48 mmHg).

These trends were generally consistent for DBP estimation, although the complete model achieved Grade C under BHS criteria, indicating lower estimation variability compared to SBP. Although removing PPG inputs slightly improved MAE from 7.77 to 7.54 mmHg, the corresponding BHS grading decreased to Grade D, indicating increased variability despite lower average error. Similar to SBP estimation, removing demographic inputs produced the largest performance degradation, increasing MAE from 7.77 to 9.96 mmHg and STD from 3.84 to 5.27 mmHg.

### 3.7 Subgroup Study Results

The subgroup analysis comparing DL model MAE between obese and non-obese cohorts is presented in Table 12, with additional visualization of all error metrics provided in Fig 11. The results demonstrate that the best-performing DL model produced higher estimation errors in obese participants compared to non-obese participants on average (SBP MAE: 10.72 vs 8.90 mmHg, DBP MAE: 13.96 vs 6.88 mmHg). For SBP estimation, this increase corresponded to a medium effect size but did not reach statistical significance ($U = 120$, $p = 0.197$, $\delta = 0.371$). For DBP estimation, the increase in MAE was statistically significant and associated with a large effect size ($U = 148$, $p = 0.011$, $\delta = 0.691$). These findings suggest that model performance degrades in participants with higher BMI, with a more pronounced effect observed for DBP estimation.

Table 12. Results of subgroup analyses on the best performing DL model for SBP and DBP estimation within obese and non-obese subgroups. Statistical significance is denoted by * for $p < 0.05$.

| | **Validation Cohort** | **MAE (mmHg)** | **ME ± STD (mmHg)** | **Mann-Whitney U (*p*-value)** | **Cliff's δ (Size Interpretation)** |
|---|---|---|---|---|---|
| **Systolic Blood Pressure (SBP)** | Obese (n = 5) | 10.72 | 0.27 ± 6.19 | 120 (0.197) | 0.371 (Medium) |
| | Non-Obese (n = 35) | 8.90 | -0.03 ± 5.52 | | |
| **Diastolic Blood Pressure (DBP)** | Obese (n = 5) | 13.96 | -1.30 ± 4.20 | 148 (0.011*) | 0.691 (Large) |
| | Non-Obese (n = 35) | 6.88 | -0.02 ± 3.78 | | |

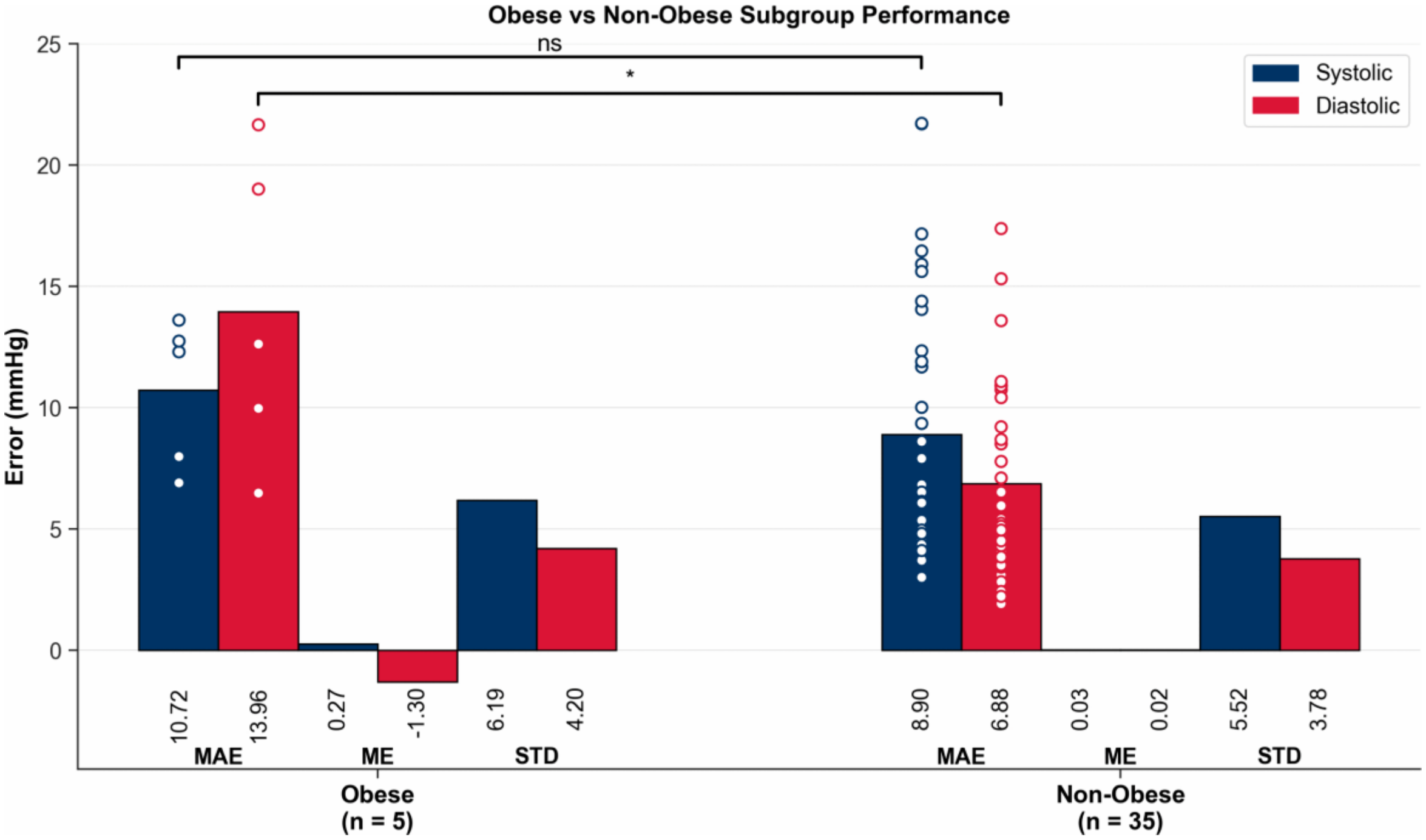

Fig 11. Grouped bar charts for comparing mean absolute error (MAE), mean error (ME), and standard deviation (STD) errors within obese and non-obese subgroups. * = $p < 0.05$, ns = not significant.

## 4 DISCUSSION

Both mechanism-based and ML models performed worse at generalizable cuffless BP estimation compared to the proposed DL model. Although several ML models achieved slightly lower MAE values than mechanism-based approaches, their substantially higher error variability limits their practical utility. This sensitivity to inter-subject variability is likely related to the relatively unconstrained prediction space of traditional ML models, which can produce inaccurate mappings between extracted features and ground truth BP values. In contrast, the proposed DL pipeline enables the model to independently learn waveform-level representations and modulate these features using demographic encodings to encourage predictions within physiologically plausible ranges for unseen subjects. Results from the baseline estimator analysis further support this workflow, suggesting that the model learns subject-specific physiological patterns rather than collapsing predictions towards the training population mean. Collectively, these findings suggest that calibration-free smartwatch-based BP estimation may be better suited for DL modelling approaches, where inter-subject variability can be more effectively minimized to support equitable performance across heterogeneous subjects.

The ablation study further supported the importance of multimodal inputs for BP estimation. Removing demographic inputs resulted in the largest increase in LOSO MAE, suggesting that these encodings are critical for shifting BP predictions toward the physiological range of unseen subjects. Given the well-established impact of age, sex, body composition, and genetics on cardiovascular health, this finding further reinforces the importance of documenting comprehensive demographic information in datasets to modulate waveform-derived predictions. The ablation study also revealed the relative contribution of each sensing modality. When combined with demographic inputs, ECG signals contributed more strongly to reducing both SBP and DBP estimation error than PPG signals. Interestingly, ECG-only models achieved lower errors for SBP estimation (MAE: 13.57 mmHg) than PPG-only models (MAE: 14.55 mmHg). Conversely, PPG-only models yielded better DBP estimation performance (MAE: 9.37 mmHg) than ECG-only models (MAE: 10.63 mmHg), suggesting that the two modalities may provide complementary information relevant to SBP and DBP estimation. Although PPG signals provide continuous monitoring utility, their susceptibility to degraded waveform morphology caused by motion, poor wrist contact, ambient lighting and other real-world factors may have limited their contribution to BP estimation in this work [52]. Smartwatch ECG signals, while also susceptible to contamination, likely provided more stable morphological information due to the distinct electrical activity associated with each cardiac cycle. Nevertheless, the fusion-based approach consistently outperformed single-modality models, with additional improvements observed following the inclusion of demographic inputs. This finding is consistent with observations from Liu et al. [33], highlighting the importance of multimodal fusion for calibration-free BP estimation using consumer-grade wearable hardware.

A limitation of these results is that the proposed model satisfied AAMI requirements for both SBP and DBP estimation but had limited success under IEEE 1708 and BHS grading standards, with the model achieving Grade D for SBP and Grade C for DBP. These findings suggest that the model exhibits low systematic bias and acceptable variability, but prediction errors do not consistently remain within clinically acceptable thresholds to reliably replace cuff-based gold standards. However, the achieved error values were comparable to several previously reported wrist-worn cuffless BP estimation studies [21, 33, 43, 48], while also being evaluated using a strict LOSO-CV framework to capture calibration-free performance on unseen subjects.

Although the proposed DL architecture achieves the strongest LOSO performance overall, degraded performance was observed for certain subgroups. Specifically, the subgroup analysis demonstrated statistically significant increases in DBP MAE for obese participants compared to non-obese participants. A similar increase was observed for SBP MAE, although this difference did not reach statistical significance. While the obese cohort size was relatively small ($n = 5$), the observed statistical significance and large effect size suggest that obesity-related physiology may meaningfully impact model performance. For example, obese individuals may exhibit increased vascular resistance and altered peripheral hemodynamics, resulting in more subtle variations in PPG waveforms that are difficult to accurately capture using smartwatch sensors. Strategies such as multi-wavelength PPG have been previously been shown to improve BP estimation compared to single-channel approaches [33]. Additionally, obesity-related differences in sweat production may affect electrodermal responses and the skin-electrode interface during wearable ECG acquisition, potentially degrading ECG signal quality compared to non-obese participants [55]. As these subgroup effects were present with the complete model, the results suggest that the demographic encodings used in this work were insufficient to fully mitigate population biases.

Additional physiological information and more advanced feature-fusion strategies may therefore be required to support more equitable performance across unseen subjects.

Several limitations should be considered when interpreting these findings. First, the relatively small dataset required segmenting each of the 200 independent recordings (40 participants x 5 measurements) into multiple windows. Because windows derived from the same recording shared a common blood pressure label, they were temporally autocorrelated rather than fully independent observations. Although windowing is a common strategy for snapshot BP studies where observations are sparse [21, 48, 71], the limited number of independent recordings may have constrained the ability of the proposed DL model to learn generalizable physiological patterns. Similarly, the small cohort size reduced statistical power for subgroup analyses, limiting statistical power for drawing population-level conclusions. Second, the demographic inputs used in this work were limited to self-reported descriptors and likely did not fully capture inter-subject physiological differences in vascular health and peripheral hemodynamics that may be beneficial for improving calibration-free performance. Third, although the variations in physical states and wearing conditions introduced semi-naturalistic variability associated with real-world use, the dataset did not capture fully free-living conditions, and the reported results should therefore not be interpreted as representative of fully unrestricted real-world deployment. In addition, the cuff-based reference BP measurements may be influenced by phenomena such as masked or white-coat hypertension, introducing variability into the ground-truth labels and potentially affecting model accuracy. Similarly, previous work has found that ECG and PPG signals can be affected by daily behaviors such as drinking coffee prior to data acquisition, which may introduce additional variability into both the BP labels and the resulting estimation accuracy [1, 26, 61]. The smartwatch ECG acquisition also requires active user interaction, limiting the practicality of continuous multimodal BP monitoring. The portability and wearable nature of smartwatches enable potential solutions to this workflow limitation, where users may transition to ECG-PPG fusion monitoring during routine or clinically relevant HBPM sessions while relying on passive PPG-based monitoring throughout the remainder of the day. From a clinical usability standpoint, converting longitudinal estimation results into categorical labels such as Normal, Elevated, and Stage 1 or 2 Hypertension could provide greater utility and interpretability for clinicians. For patients with at-risk BP patterns or undergoing treatment regimes, the underlying estimations from the cuffless BP algorithms could be then examined further to support clinical monitoring and decision-making.

The findings of this work demonstrate the potential of consumer-grade smartwatch hardware for multimodal cuffless BP estimation using calibration-free approaches. Future work should investigate how larger and more diverse cohorts, free-living conditions, additional sensing modalities, and advanced fusion strategies to improve generalizability to unseen subjects. Although consumer-grade smartwatches offer a promising and accessible platform for longitudinal HBPM algorithms, further improvements are required before reliable deployment under real-world conditions can be achieved.

**ACKNOWLEDGMENTS**

The authors would like to thank the Ontario Graduate Scholarship for supporting this research.